\documentclass[aps,prl,twocolumn,final,letterpaper]{revtex4}

\usepackage{graphicx}   
\usepackage{import}                         
\usepackage{epstopdf}
\usepackage{amsmath} 
\usepackage{bm}
\usepackage{amssymb}
\usepackage{quotes}
\usepackage{transparent}
\usepackage{dcolumn}
\usepackage[usenames, dvipsnames]{color}
\usepackage{multirow}
\usepackage{cancel} 
\usepackage{mdframed}
\usepackage{color}
\usepackage{bm}
\usepackage{setspace}
\usepackage{dsfont}
\usepackage{braket}
\usepackage{slashed}
\usepackage{float}
\usepackage{soul, color} 
\soulregister\ref{7}  
\soulregister\cite{7} 
\usepackage{physics}

\begin{document}
\rmfamily

\title{Quantum Electrodynamical Metamaterials}
\author{Josephine Yu$^{1*}$, Jamison Sloan$^{3*}$, Nicholas Rivera$^{2}$, and Marin Solja\v{c}i\'{c}$^{2}$}

\affiliation{$^{1}$ Department of Applied Physics, Stanford University, Stanford, CA 94305, USA\\
$^{2}$ Department of Physics, MIT, Cambridge, MA 02139, USA \looseness=-1}
\affiliation{$^{3}$ Research Laboratory of Electronics, MIT, Cambridge, MA 02139, USA \looseness=-1\\
$^{*}$These authors contributed equally.}

\noindent	

\begin{abstract}

Recent experiments have revealed ultrastrong coupling between light and matter as a promising avenue for modifying material properties, such as electrical transport, chemical reaction rates, and even superconductivity. Here, we explore (ultra)strong coupling as a means for manipulating the optical response of metamaterials based on ensembles of constituent units individually in the ultrastrong coupling regime. We develop a framework based on linear response for quantum electrodynamical systems to study how light-matter coupling affects the optical response. We begin by applying this framework to find the optical response of a two-level emitter coupled to a single cavity mode, which could be seen as a ``meta-atom'' of a metamaterial built from repeated units of this system. We find optical behaviors ranging from that of a simple two-level system (Lorentz-oscillator) to effectively transparent, as the coupling goes from the weak to deep strong coupling regimes. We explore a one-dimensional chain of these meta-atoms, demonstrating the tunability of its optical behavior. Our scheme may ultimately provide a framework for designing new metamaterials with low-loss, highly-confined modes, as well as tunable (single-photon) nonlinearities. 
\end{abstract}

\maketitle

\noindent

The interactions between light and matter form the basis for describing, and ultimately controlling, the optical response of materials. The optical response of a material is governed by its microscopic dynamics at a quantum mechanical level, often understood through the energy eigenstates of electrons, phonons, or other excitations of the system. Typically, descriptions of a material's optical properties rely on a quantum mechanical description of the constituent matter of the system, but only a classical description of the electromagnetic field. This is usually sufficient, because in most cases, the constituent particles of a material are only weakly coupled to the surrounding electromagnetic field modes.

However, recent experiments in strong and ultrastrong coupling have shown that the the quantized nature of the electromagnetic field can strongly influence material properties. In fact, from its first experimental observation in 2009 \cite{anappara_signatures_2009}, ultrastrong coupling has revealed itself as a promising avenue for controlling various material properties \cite{ashhab_qubit-oscillator_2010, galego_cavity-induced_2015,orgiu_conductivity_2015,  herrera_cavity-controlled_2016, kockum_deterministic_2017}, including chemical reaction rates \cite{hutchison_modifying_2012}, superconducting temperatures \cite{thomas_exploring_2019}, and magnetotransport in 2D electron gases \cite{paravicini-bagliani_magneto-transport_2019}. Moreover, ultrastrong coupling is a ubiquitous phenomenon, observed in a variety of experimental platforms ranging from those involving collective excitations of emitters \cite{gambino_exploring_2014, todisco_ultrastrong_2018, mazzeo_ultrastrong_2014, gunter_sub-cycle_2009, delteil_charge-induced_2012, auer_entangled_2012}, to systems such as superconducting circuits \cite{niemczyk_circuit_2010, forn-diaz_observation_2010} which demonstrate ultrastrong coupling between a single two-level emitter (qubit) and light. Recent experiments using superconducting circuits have also demonstrated ultrastrong coupling between artificial atoms and a continuum of electromagnetic modes \cite{forn-diaz_ultrastrong_2017, puertas_martinez_tunable_2019, magazzu_probing_2018}, further widening the scope of possibilities for ultrastrong coupling. 

Despite the promise of these recent developments, strong and ultrastrong coupling have not yet been considered as a platform on which to design new macroscopic-scale optical metamaterials. This is a particularly intriguing prospect, since control over optical materials may present opportunities for optical excitations with low loss and high confinement, tunable nonlinearities, and at frequency ranges which are naturally difficult to achieve. Here, we propose using ultrastrong coupling between light and matter to create novel optical materials with tunable properties. We show that even a two-level system coupled to a single electromagnetic mode, which we refer to as a ``USC meta-atom,'' or just ``meta-atom,'' exhibits a rich variety of optical behavior depending on the coupling strength between the emitter and cavity, ranging from the well-known behavior of a two-level system (Lorentz oscillator) to effectively transparent. 
We propose to design new metamaterials based on meta-atoms and show that their macroscopic optical response can be reshaped by tuning the coupling strength between the light and matter in the constituent meta-atoms of the metamaterial; by extension, we also control the new electromagnetic modes which propagate through such a metamaterial. We discuss how the quantum electrodynamical optical materials we propose could be realistically implemented via systems such as superconducting qubits \cite{niemczyk_circuit_2010} or molecules in picocavities \cite{chikkaraddy_single-molecule_2016}, as these systems have achieved the requisite USC strengths and could be arranged into a metamaterial with straightforward geometry, such as a one- or two-dimensional array. 

\begin{figure}
\begin{center}
\includegraphics[scale=1]{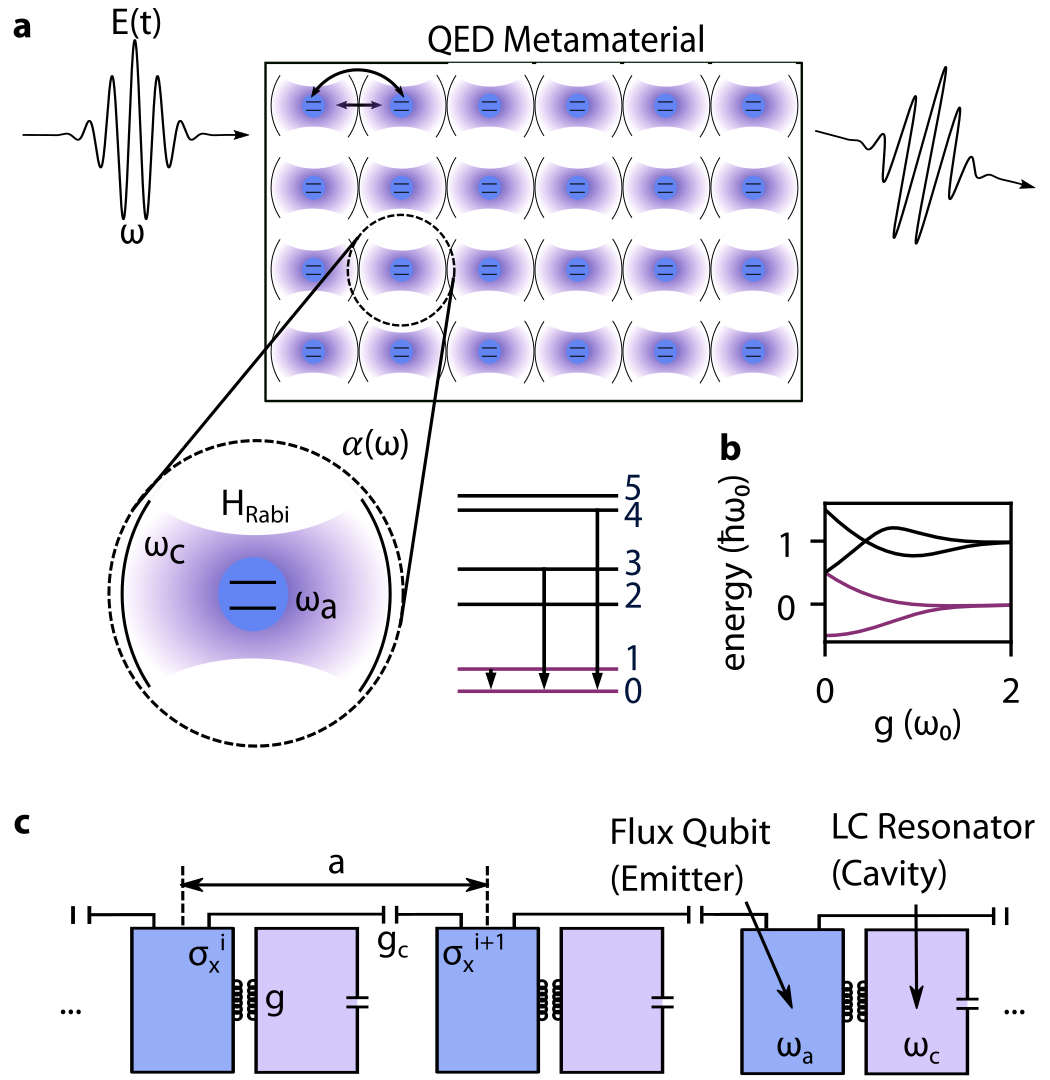}
\end{center}
\caption{(a) Schematic of the proposed ``strongly coupled metamaterial'' and interacting field described by $E(t)$ with driving frequency $\omega$. The meta-atoms comprising the metamaterial can interact via dipole or cavity coupling, or via a more general interaction involving both the dipole and cavity. An energy level diagram for the meta-atom shows an example of the allowed transitions at a particular coupling strength. (b) A plot displaying how the first few energy levels evolve as a function of $g$. As the coupling increases, the two lowest energy levels converge, bringing the lowest resonant transition energy to zero. (c) Schematic of the proposed physical realization of a 1D material composed of meta-atoms. Each meta-atom consists of a flux qubit inductively coupled to an LC resonator with an effective coupling $g$. Neighboring units are coupled capacitively through their charges, generating an inter-unit coupling $g_c \sigma_x^i \sigma_x^{i+1}$ between the $i$th and $i+1$th qubits.
\label{fig:schematic}}
\end{figure}

We now elaborate on our proposed model for ``strongly coupled metamaterials'' which are composed of individual strongly coupled ``meta-atoms'' as their constituent components. Each meta-atom is comprised of a two-level system with frequency $\omega_a$ coupled to a single electromagnetic mode of frequency $\omega_c$, represented as a resonant cavity. This model of a meta-atom serves as a versatile proxy for physical systems in which ultrastrong coupling has been achieved. We also note that while we focus on the case of a two level system coupled to a single mode, the concept of a meta-atom is quite general, and encompasses possibilities such as multi-level emitters (molecules, artificial atoms, spins, etc.) and multi-mode electromagnetic excitations (fiber modes, photonic crystal modes, surface polaritons, etc.). Once the platform has been chosen, a metamaterial can be conceived as a sufficiently dense arrangement of many meta-atoms, shown in Fig. \ref{fig:schematic}a. 

To provide a thorough analysis of an example of the proposed material, we consider one possible realization of the meta-atoms: flux qubits, functioning as the two-level systems, coupled to LC resonators, functioning as the cavities. We will later focus on a chain of such units and explore the modes supported in such a one-dimensional material. In our example, the meta-atoms interact through their dipoles (Fig. \ref{fig:schematic}c), but interactions between meta-atoms can in general involve both the cavity and two-level system. 

We begin by describing the optical response of an individual meta-atom. We assume that each meta-atom is a two-level system coupled to a single cavity mode, and is thus described by the Rabi Hamiltonian 
\begin{equation}
    H_{\text{Rabi}} = \frac{\hbar \omega_a}{2} \sigma_z + \hbar \omega_c \hat{a}_c^\dagger \hat{a}_c + \hbar g  \sigma_x \left(\hat{a}_c^\dagger  + \hat{a}_c \right) + \frac{\hbar g^2}{\omega_c},
    \label{eq:rabi_bare}
\end{equation}
where $\omega_a$ is the transition frequency of the two-level system, $\omega_c$ is the cavity frequency, $\hat{a}^{(\dagger)}$ is the annihilation (creation) operator for the cavity mode, $g$ is the light-matter coupling strength, and $\sigma_x$ and $\sigma_z$ are Pauli matrices. We consider here the case that the emitter and cavity are on resonance at a frequency $\omega_0$ such that $\omega_a = \omega_c = \omega_0$. As the emitter-cavity coupling strength $g$ changes, so do the energy levels and allowed transitions in the meta-atom, as highlighted in Fig. \ref{fig:schematic}b. The nature of the interactions between light and matter in a system is characterized by comparing $g$ to the losses or transition energies of the system \cite{frisk_kockum_ultrastrong_2019}. When the coupling strength is lower than the losses in the system, the system is in the weak coupling (WC) regime. Here, interactions may be treated perturbatively, resulting in relatively well-studied phenomena in which the eigenstates of the emitter are not drastically impacted by the coupling to electromagnetic modes, such as Purcell enhancement of spontaneous emission \cite{purcell_spontaneous_1995}. In contrast, the strong-coupling (SC) regime is defined by a coupling strength $g$ which exceeds the losses in the system. Physically, the matter and photon form a single entity whose energy levels differ substantially from those of the individual components, giving rise to phenomena such as Rabi oscillations. As the coupling increases to the order of the transition energies, the system enters the ultrastrong coupling (USC) regime, in which even properties of the quantum-mechanical ground state are strongly modified \cite{ashhab_qubit-oscillator_2010, hines_entanglement_2004}.

To compute the optical response of the meta-atom, we introduce a classical probe-field of frequency $\omega$ through a time-dependent perturbation $E(t)$. We assume that this probe-field couples to the dipole moment $d$ of the two-level system in the meta-atom. While more general driving mechanisms exist, the direct drive of the two-level system is readily realizable in the flux qubit systems that we use as an example here. Then, taking the perturbation to be sinusoidal, the full Hamiltonian takes the form
\begin{align}
 H = H_{\text{Rabi}} + dE_0 \sigma_x\left(e^{i\omega t} + e^{-i\omega  t}\right).
\label{eq:rabi}
\end{align}
For a system like the proposed flux qubit (two-level system) coupled to an LC resonator (cavity), the natural observable quantities are those involving the transitions between eigenstates of the Rabi Hamiltonian (Rabi basis states). Since we aim to characterize the response of the meta-atom, rather than the emitter alone, we should analyze the expectation values of the ``effective" dipole operators corresponding to each transition between Rabi basis states $m$ and $n$: $\tau_x^{nm}\equiv \ket{n}\bra{m}+\ket{m}\bra{n}$. 

We now consider the case that the driving frequency $\omega$ in Eq. \ref{eq:rabi} is close to a transition frequency $\omega_{n0}$ of the meta-atom between the $n$-th excited state of the Rabi Hamiltonian and the ground state. Then, it is possible to relate the emitter's dipole operator $\sigma_x$ and the effective dipole operator for the meta-atom $\tau_x^{n0}$.  By expanding the emitter's dipole operator $\sigma_x$ explicitly in the Rabi basis, we find $\sigma_x = \sum_{m,n} \sigma_x^{mn}\ket{m}\bra{n}$. When $\omega \approx \omega_{n0}$, the system is well-approximated by a two-level system with levels $\ket{0}$ and $\ket{n}$. In this case, the terms $\sigma_x^{n0}\ket{n}\bra{0}$ and $\sigma_x^{0n}\ket{0}\bra{n}$ contribute most in the expansion of $\sigma$, so we can approximate the dipole operator operator $\tau_x^{n0}$ in terms of the emitter's effective dipole operator, 
\begin{equation}
    \tau_x^{n0} = \frac{\sigma_x}{\sigma_x^{0n}}, \label{eq:taudef}
\end{equation}
where $O^{mn} \equiv \braket{m|O}{n}$ denotes matrix elements between eigenstates of the coupled system. The full meta-atom is then approximately a collection of two-level systems with transition frequencies which can be tuned via the emitter-cavity coupling $g$. 

\begin{figure*}
\begin{center}
\includegraphics[scale=1]{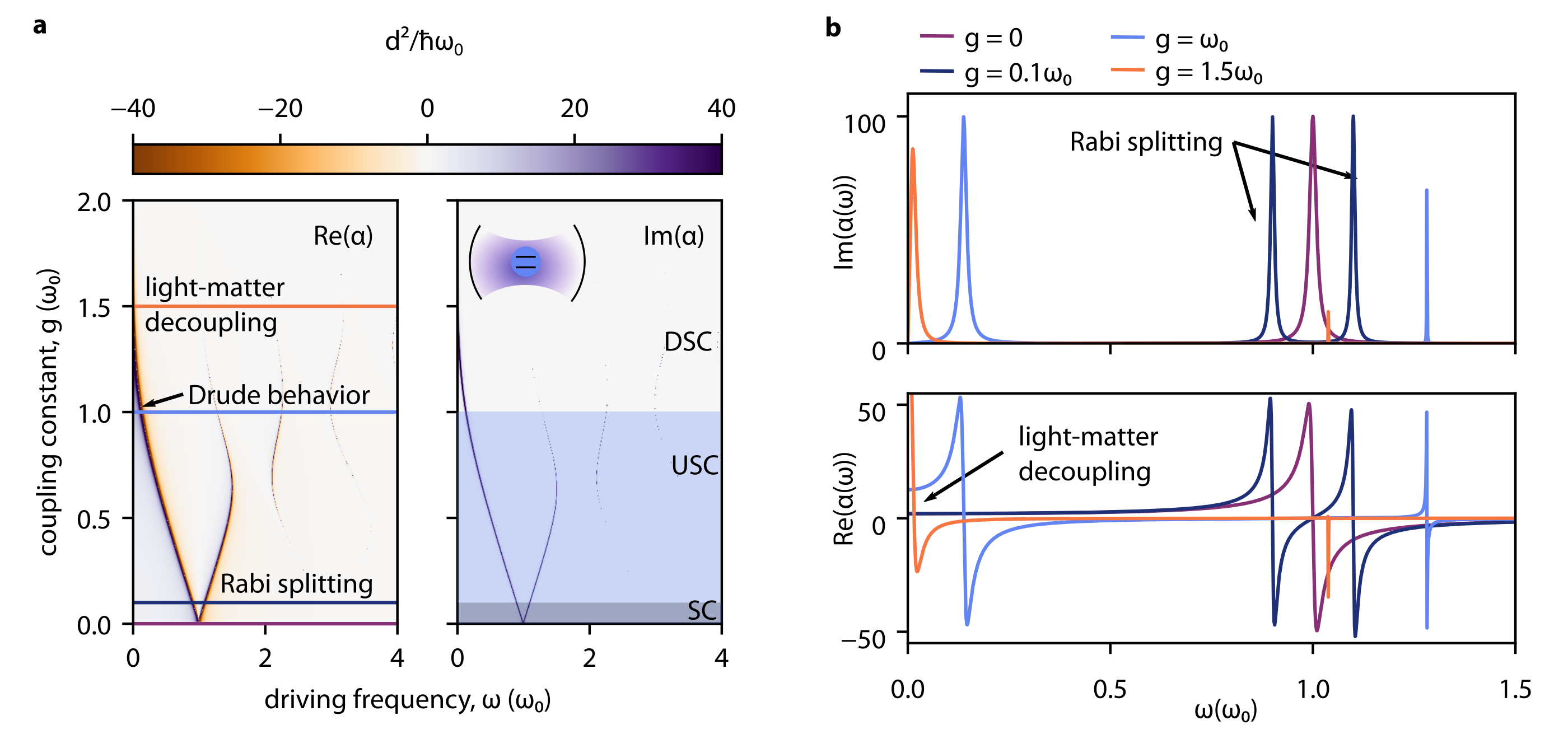}
\caption{(a) Atomic polarizability of the coupled atom-cavity system with phenomenological loss $\gamma_a = 0.01\omega_0$ (and omitting cavity loss $\gamma_c$) as a function of driving frequency $\omega$ and coupling strength $g$. Colored lines at $g=0$  $g=0.1\omega_0$, $g=\omega_0$, and $g=1.5\omega_0$  highlight signature behaviors at various coupling strengths, which correspond to the various regimes of light-matter interaction. (b) Polarizability of a single meta-atom at various values of the coupling strength $g$ highlighted in the line cuts of (a). \label{fig:pol_color}}
\end{center}
\end{figure*}


Though the Hamiltonian in Eq. \ref{eq:rabi} is valid for arbitrary drive strengths, we focus on weak probe-fields so that we may treat the probe-field interaction $E(t)$ as a small perturbation on the meta-atom Hamiltonian $H_{\text{Rabi}}$. The optical response of the meta-atom is then well-captured by the first-order correction of the effective dipole expectation value (Eq. \ref{eq:taudef}) in response to the applied external field. This correction is proportional to the atomic polarizability $\alpha(\omega)$, a linear response function which relates the applied electric field $E(\omega)$ to the change in dipole moment $\langle d(\omega) \rangle = \alpha(\omega)E(\omega)$ in frequency space.
Using time-dependent perturbation theory to find the first-order correction to the dipole expectation value, we find that the frequency-dependent polarizability $\alpha(\omega)$ for a strongly coupled system initially in the ground state can be approximated as the sum of responses from all excited levels:
\begin{align}
\label{eq:polarizability}
\begin{split}
\alpha(\omega) &= \frac{d^2}{\hbar}\sum_n \alpha_{n0}(\omega),
\\
\alpha_{n0}(\omega) &= -I_{n0}  \left[\frac{1}{\omega-\omega_{n0} - i\Gamma_n} - \frac{1}{\omega+\omega_{n0} - i\Gamma_n} \right].
\end{split}
\end{align}
Here, $\omega_{n0}$ is the transition frequency in the coupled system,  $I_{n0}$ is 0 if $\braket{n|\sigma_x}{0} = 0$ and 1 otherwise, and $\Gamma_n$ is the loss associated with the transition from the $n$-th excited state to the ground state of the coupled system. This result from perturbation theory has been corroborated by a numerical approach via master equation evolution. 

In the weak coupling regime, the linewidth $\Gamma_n$ is given simply by the atomic linewidth of the two-level system transition. However, the onset of the ultrastrong coupling regime invalidates this assumption. Since the ground state of an ultrastrongly coupled system contains virtual excitations, naively relaxing the system according to the losses and transitions of the emitter would result in unphysical photon generation in the cavity. In Eq. \ref{eq:polarizability}, we account for the effects of (ultra)strong coupling and define the loss associated with each transition according to \cite{beaudoin_dissipation_2011} as $\Gamma_n \equiv |\braket{n| \sigma_x}{0}|^2\gamma_a + |\braket{n| (\hat{a} + \hat{a}^\dagger)}{0}|^2\gamma_c $, where $\gamma_a$ and $\gamma_c$ are the relaxation rates of the emitter and cavity in the absence of coupling, respectively. We assume for concreteness that the atomic relaxation rate dominates the overall loss and therefore neglect the cavity relaxation rate. Qualitatively, the inclusion of the cavity loss would not change our results. Additionally, we assume that the atomic relaxation rate is substantially smaller than the differences between transition frequencies in the system, ensuring that the transition frequencies of the coupled system remain distinct. This assumption ensures that the relaxation due to losses in the emitter and the cavity may be treated independently (see \cite{beaudoin_dissipation_2011} for details). We note that for the atomic relaxation rate of $\gamma_a = 0.02\omega_0$ used throughout our calculations, the previously stated assumptions are valid up until a coupling strength of $g=1.5\omega_0$, at which the lowest transition frequency approaches the value of the $\gamma_a$: $\omega_{10} \sim 0.01\omega_0$. 

The form of Eq. \ref{eq:polarizability} shows how the optical response of a single meta-atom -- namely, the location, height, and width of each peak -- is strongly controlled by the coupling. Both the eigenstates (which contribute to the matrix elements) and the eigen-energies (which contribute to the transition frequencies, as seen in Fig. \ref{fig:schematic}b) evolve as a function of the coupling constant $g$. In Fig. \ref{fig:pol_color}a, the changes in transition energies are reflected in the evolving location of the peaks in the atomic polarizability, as well as in the introduction of new peaks at sufficiently high coupling strengths. The line cuts at various coupling strengths (Fig. \ref{fig:pol_color}b) highlight features in the meta-atom system representative of the corresponding regimes of light-matter interaction. In the WC regime ($g \ll \gamma_a$), the system unsurprisingly behaves as a single Lorentz oscillator with resonant frequency $\omega_0$, since the atomic energy structure is completely unaffected by the presence of the vacuum electromagnetic field. As the coupling strength increases ($g=0.1\omega_0$), the polarizability exhibits a signature of the strong coupling regime: the single peak splits into two, indicative of Rabi splitting. Under strong coupling ($g \sim \gamma_a$), the system, which previously had just one resonant transition energy, now has an expanded set of energy levels, causing the previously degenerate excited states to split into distinct upper and lower polariton states, as can be described by the Jaynes-Cummings (JC) Hamiltonian \cite{jaynes_comparison_1963}. Once the system enters the ultra-strong coupling regime ($g = \omega_0$), the JC model breaks down, and new resonant frequencies are introduced as a result of the inclusion of counter-rotating terms in the Rabi Hamiltonian (Eq. \ref{eq:rabi}). This results in the appearance of multiple new peaks between $g=0.1\omega_0$ and $g=\omega_0$ in Fig. \ref{fig:pol_color}b. The introduction of new resonant frequencies originates physically from the mixing between light and matter states, which expands the number of possible transitions (and therefore transition frequencies) as the coupling increases. The line cut at $g=\omega_0$ also highlights a region in which the lowest-frequency peak in the response approaches $\omega=0$. This occurs as the ground and first excited states of the coupled system converge (see Fig. \ref{fig:schematic}b), thus causing the corresponding transition energy to approach zero. 

Beyond dictating the number and location of resonances (and therefore peaks) in the system, the coupling strength $g$ also plays a key role in determining the losses associated with each transition, manifested through the widths and heights of the corresponding peaks. The evolving eigenstates dictate the matrix elements of $\sigma_x$ by which the loss parameter $\Gamma_n$ is scaled; as the coupling $g$ increases, these matrix elements approach zero in all transitions except for the lowest, thus bringing the scaled loss $\Gamma_n$ to zero for $n\neq 1$ and thereby narrowing the corresponding peaks of the transitions from the $n-$th excited level to the ground state. In Fig. \ref{fig:pol_color}b, the narrowing of peaks can be seen especially clearly in regions of higher $g$ in the colorplots. Indeed, the line cut at $g=1.5\omega_0$ in Fig. \ref{fig:pol_color}b highlights how the polarizability resembles that of a transparent emitter which does not interact with the external probe at nonzero frequency and sufficiently high coupling strengths. 

So far, we have shown that even the simplest possible strongly coupled light-matter system --- a two-level system coupled to a single mode --- exhibits a rich variety of optical behavior depending on the coupling strength. The optical response of a material is often dictated by the quantum mechanical properties of the constituent particles, so the tunable and varied behaviors of the meta-atom make it promising foundation for designing similarly behaved bulk materials. We now focus on one example of the individual meta-atom properties translating to a material. Here, we show that even a simple 1D chain of meta-atoms inherits their strong- and ultrastrong-coupling behaviors to give rise to novel modes. 

To be concrete, we consider a 1D chain of the emitter-resonator units realized for example through flux qubits inductively coupled to LC resonators (Fig. \ref{fig:schematic}c). Units are capacitively coupled to only their nearest neighbors through charge interactions, which may be expressed as $g_c \sigma_x^i \sigma_x^j$ between the qubits at neighboring sites $i$ and $j$ \cite{krantz_quantum_2019}. Here, the coupling constant $g_c$ is determined by the circuit elements (i.e., the capacitance within each qubit and the coupling capacitance). Since the only coupling is facilitated through macroscopic wires, it is only the charges across different meta-atom units that interact with each other. We assume that the capacitances involved are frequency-independent in the range of interest, and we set the coupling constant to take the value $g_c = 0.08\omega_0$ \cite{ozfidan_demonstration_2020, yan_flux_2016, krantz_quantum_2019}. Since the coupling between meta-atoms is weak compared to all other energy scales, we expect that the effect of coupling is perturbative and that the supported modes are close in frequency to the resonant frequencies of the individual  units. Near each resonant frequency of the meta-atom, we may express the interaction in terms of the transition operators in the Rabi basis (Eq. \ref{eq:taudef}) as $g_c |\sigma^{0n}|^2 \tau_x^i \tau_x^j$. One can interpret this as a dipole $\tau_x^j$ interacting with a field $g_c |\sigma^{0n}|^2 \tau_x^i$ generated by its neighbor.
 
 \begin{figure}
\begin{center}
\includegraphics[scale=1]{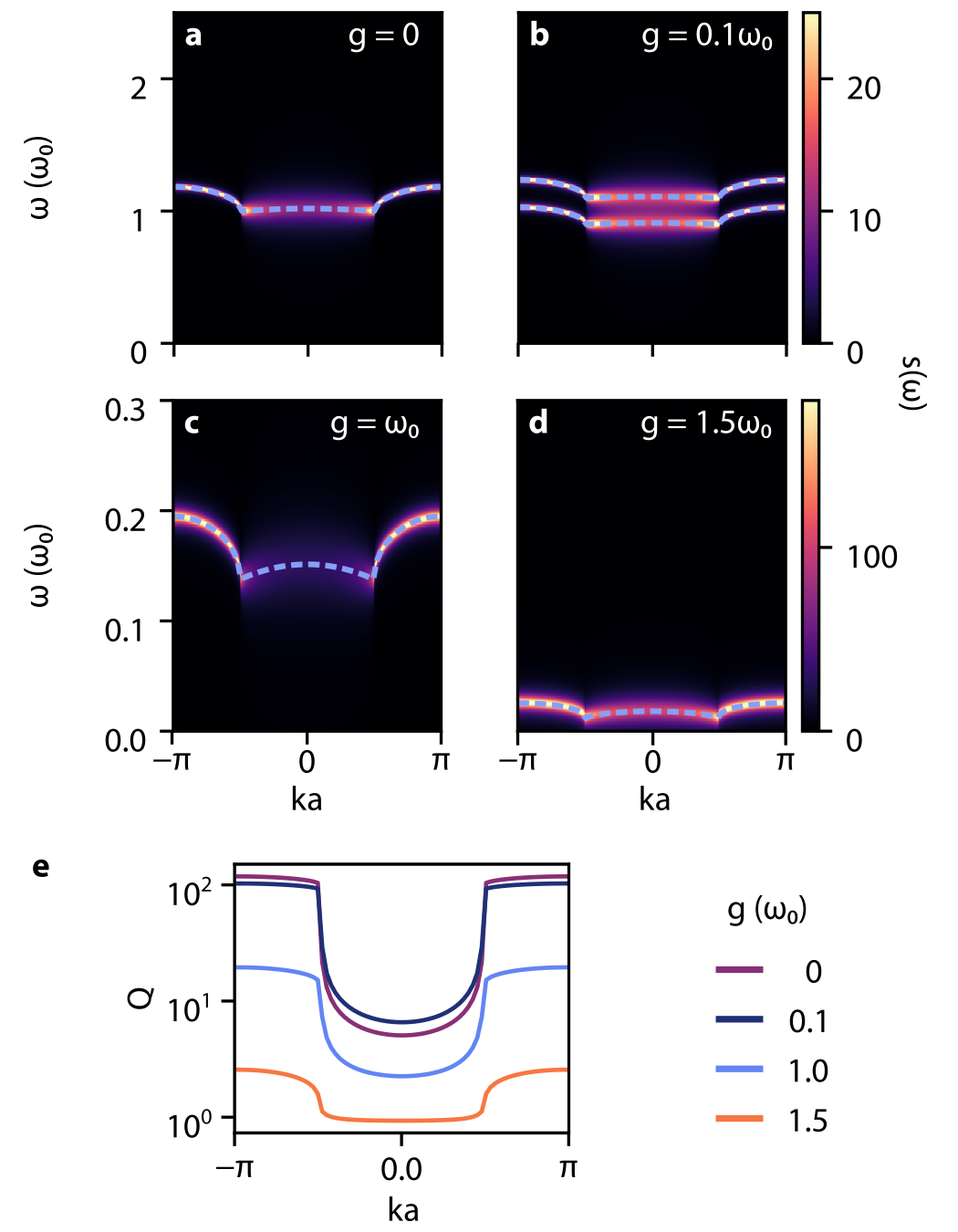}
\end{center}
\caption{(a)-(d) Dispersion relations for a one-dimensional chain of meta-atoms for various coupling values $g$, illustrated through a colormap of the spectral function $s(k, \omega)$. Overlaid is Re$[\omega(k)]$. (e) Quality factor for the lowest-frequency mode as a function of $ka$, for various coupling values $g$.  \label{fig:chaindisp}}
\end{figure}
 
Each unit has an effective polarizability as calculated in Eq. \ref{eq:polarizability}, so we model the geometry as a line of polarizable units and calculate the supported modes by solving the equation of motion for the self-consistent local dipole moment $\tau_x$ of each unit. In a spatially periodic system, $\tau_x^i(t) = e^{ikx}\Tilde{\tau}_x^i (t)$, where $\Tilde{\tau}_x$ is the envelope of the periodic function and has no spatial dependence. Without any inter-unit coupling, the effective dipole moment oscillates at any of its resonant frequencies, $\omega_{n0}$. The dipole moment is also affected by the fields from its neighbors. In full, the equation of motion for the $i$-th qubit's dipole moment is:
\begin{equation}
    -\omega^2 \Tilde{\tau}_x^i = \omega_{n0}^2(\Tilde{\tau}_x^i +\alpha_{n0}(\omega) g_c |\sigma^{0n}|^2 (\Tilde{\tau}_x^{i-1}e^{-ika} +\Tilde{\tau}_x^{i+1}e^{ika})), 
\label{eq:eom_tau}
\end{equation}
where $a$ is the lattice spacing, and $k$ is the wavevector of the propagating mode. Since $\Tilde{\tau}_x^i$ has no spatial dependence, Eq. \ref{eq:eom_tau} may be simplified and solved to find the supported modes of frequency 
\begin{equation}
    \omega = \omega_{n0}\sqrt{1+ \alpha_{n0}(\omega) |\sigma^{0n}|^2 g_{c} \cos ka }. \label{eq:1Dmode}
\end{equation}

We now describe the properties of the modes described by Eq. \ref{eq:1Dmode}. We solve for the form of the dispersion relation $\omega(k)$ via both an analytical approximation and numerical methods. Analytically, we approximate $\alpha_{n0}(\omega)$ by the term $1/(\omega-\omega_{n0}-i\Gamma_n)$, which dominates near resonance. The resulting cubic equation can be solved numerically, yielding only one solution which respects passivity (i.e., it decays over time). We focus here on that solution.  

Fig. \ref{fig:chaindisp}b-e shows the exact dispersion relations (found numerically from Eq. \ref{eq:1Dmode}) for the allowed modes of the chain for various coupling strengths $g$ between the emitter and resonator of each unit. Like in the case of a single unit, the allowed modes at a given emitter-cavity coupling value $g$ are dictated by the resonant frequencies of the system. As expected, the chain supports a mode only near the transition frequency of the emitter if there is no emitter-resonator coupling ($g=0$). Small values of $g$ result in Rabi splitting ($g=0.1\omega_0$) of the single mode into two, mirroring the behavior of the energies of the individual meta-atoms.  The weak coupling behavior of this system is qualitatively similar to that which is observed in photonic crystals \cite{Chong2007}. Meanwhile, larger coupling values reveal evidence of light-matter decoupling ($g=1.5\omega_0)$, corresponding to the lowest transition frequency of the individual units being driven to zero with increasing coupling. The behaviors of the chain resemble those of the constituent meta-atoms, inheriting the signatures of strong and ultrastrong coupling behavior. 

Near $|ka|=\pi/2$, the change in sign of $\cos ka$ creates a cusp in the real part of the frequency. This is most evident from the cubic approximation used to solve Eq. \ref{eq:1Dmode}, which can be further approximated as a quadratic by neglecting the unphysical solution near $-\omega_{n0}$. The resulting quadratic equation is solved by 
\begin{equation}
    \omega(k)_{\pm} = \omega_{n0} -\frac{i\Gamma_n}{2} \pm i \sqrt{2\omega_{n0}g_c|\sigma^{n0}|^2\cos ka + \Gamma_n^2}. \label{eq:omegapm}
\end{equation}
For a passive system such as this one, the only physical solutions are those which decay in time, and in this case only $\omega_{-}(k)$ satisfies this condition. Since the coefficient of $\cos ka$ is much larger than $\Gamma_n^2$, the behavior of the mode on either side of $|ka|=\pi/2$ is dictated by the sign of $\cos ka$. For $|ka|<\pi/2$, Re$[\omega_{-}(k)]  \approx \omega_{n0}$ while for $|ka|>\pi/2$,  Re$[\omega_{-}(k)]  \approx \omega_{n0} + \sqrt{2\omega_{n0}g_c|\sigma^{n0}|^2 \cos ka}$. This sharp change in behavior is present across all coupling values and is a consequence of the geometry, coupling mechanism, and the constituent meta-atoms' optical properties. 

The modes supported by the chain structure may be further understood by examining their associated losses. For this purpose, we define the spectral function $s(k,\omega)$ which describes the relative modal loss, 
\begin{equation}
    s(k,\omega) \equiv \frac{\omega''(k)/2 }{\omega''(k)^2/4 + (\omega - \omega'(k))^2},
\end{equation}
where the frequency is decomposed into real and imaginary parts as $\omega(k) = \omega'(k) + i\omega''(k)$. A colormap of $s(k,\omega)$ is shown in Fig. \ref{fig:chaindisp}b-e. For all values of the coupling, the loss associated with modes for which $|ka|>\pi/2$ is significantly less than the loss for those with $|ka|<\pi/2$, as seen by the brighter halos around the bands for $|ka|>\pi/2$. This can be understood by examining the solution $\omega_-(k)$ in Eq. \ref{eq:omegapm}; for $|ka|>\pi/2$, there is a constant loss of magnitude $\Gamma_n/2$ as opposed to a loss of $\Gamma_n/2+\sqrt{2\omega_{n0}g_c|\sigma^{n0}|^2\cos ka}$ for $|ka|<\pi/2$.  For small values of $g$, $\omega_{n0}|\sigma^{n0}|^2$ is largest (Fig, \ref{fig:chaindisp}a), so the difference between the behaviors of the mode in regions $|ka|>\pi/2$ and $|ka|<\pi/2$ is the clearest. As the emitter-cavity coupling within each meta-atom increases, the loss associated with the mode in the region $|ka|<\pi/2$ decreases. This is due to the lowest transition frequency $\omega_{10}$ approaching zero, causing the loss in the region $|ka|<\pi/2$ to be dramatically reduced (Fig. \ref{fig:chaindisp}e). 

The quality factor (Fig. \ref{fig:chaindisp}f) is another metric by which to gauge the modal losses as functions of both the coupling $g$ and the wavevector $k$. Examining $Q \equiv \omega'(k)/\omega''(k)$, we see that the quality factor of the lowest-frequency mode is significantly higher in the region $|ka|>\pi/2$ than in the region $|ka|<\pi/2$. This behavior is also seen in Fig.\ref{fig:chaindisp}b-e, in which the losses for $|ka|>\pi/2$ are smaller than for $|ka|<\pi/2$. As the coupling $g$ increases, the quality factor actually decreases even though the $\omega''(k)$ is also decreasing; this is due to $\omega'(k)$ decreasing linearly with $\omega_{n0}$ to zero, while $\omega''(k)$ decreases like $\sqrt{\omega_{n0}}$. Additionally, the constant loss $\Gamma_n$ also increases, approaching $\gamma_a$ as $g$ increases, since the matrix element $|\sigma_x^{01}|^2$ approaches 1 with increasing $g$. The overall quality factor then decreases with increasing coupling. 

The properties of the modes sustained in this 1D chain of meta-atoms are closely related to those of the constituent meta-atoms. By tuning the emitter-cavity coupling within each meta-atom unit, one can manipulate the resonant frequencies, losses, and band structures of the resulting material excitations. As demonstrated through this example, the mode properties depend strongly not only on the meta-atom responses but also on the geometry and coupling mechanisms. 

In summary, we have presented a framework to design novel optical materials based on quantum electrodynamical strong coupling. This framework shows quite clearly the important role played by the vacuum field in determining optical properties outside of the weak coupling regime. We described the optical response of the simplest possible light-matter system --- a meta-atom composed of a coupled cavity mode and two-level emitter --- and found that it depends strongly on the coupling between the emitter and cavity. Notably, we found that by adjusting the coupling, we can change the optical behavior in a unit of this type from that of a Lorentz oscillator at a given frequency to effectively transparent. We were then able to create and control new electromagnetic excitations in an example material built of meta-atoms which inherits similar properties from its constituents. Specifically, we showed that the modes propagating through a 1D geometry of the meta-atoms inherit similar tunable properties from the consitutent meta-atoms, with modes ranging from single-emitter type responses to effectively transparent. The interplay between the geometry and the optical properties of the meta-atoms result in unique behaviors for the modes supported by the 1D material, as seen in the varied behavior of the modes across different wavenumbers $k$. 

Though we focus here on a material built of single-emitters coupled to single electromagnetic modes, our methods may also be extended to study the optical behaviors of materials in which (ultra)strong coupling occurs between more general collections of emitters and electromagnetic excitations. As a result, our work may also be related to recent experimental observations such as modified conductivity due to ultrastrong coupling \cite{orgiu_conductivity_2015}. Our approach may also be used to study the optical properties of composite systems in which USC has been proposed, such as a material formed by interfacing 2D and 3D materials \cite{kurman_tunable_2020}.


Additionally, while our work focuses on linear optical response, our methods may be extended to study nonlinear response in (ultra)strongly coupled materials, which is recent topic of interest \cite{kockum_deterministic_2017, garziano_multiphoton_2015, kockum_frequency_2017}. One potential approach is to consider higher-order terms in perturbation theory in our calculation of the dipole expectation, which would offer finer corrections to the calculations presented here in the linear regime. This approach could be used to study effects such as frequency conversion in strongly coupled materials. We can also generalize our results to nonperturbative regimes by using stronger probe fields, thus allowing the study of phenomena such as high harmonic generation in strongly coupled materials. Moreover, while we have focused on classical light in strongly coupled systems, one could also consider how these systems interact with light at a single photon level, presenting opportunities for quantum optics.

\section{Acknowledgements}
JY was supported by the National Science Foundation Graduate Research Fellowship under Grant No. DGE-1656518. This material is also based upon work supported in part by the
U. S. Army Research Office through the Institute for Soldier Nanotechnologies at MIT,
under Collaborative Agreement Number W911NF-18-2-0048.


\bibliography{references}

\begin{thebibliography}{33}
\expandafter\ifx\csname natexlab\endcsname\relax\def\natexlab#1{#1}\fi
\expandafter\ifx\csname bibnamefont\endcsname\relax
  \def\bibnamefont#1{#1}\fi
\expandafter\ifx\csname bibfnamefont\endcsname\relax
  \def\bibfnamefont#1{#1}\fi
\expandafter\ifx\csname citenamefont\endcsname\relax
  \def\citenamefont#1{#1}\fi
\expandafter\ifx\csname url\endcsname\relax
  \def\url#1{\texttt{#1}}\fi
\expandafter\ifx\csname urlprefix\endcsname\relax\def\urlprefix{URL }\fi
\providecommand{\bibinfo}[2]{#2}
\providecommand{\eprint}[2][]{\url{#2}}

\bibitem[{\citenamefont{Anappara et~al.}(2009)\citenamefont{Anappara,
  De~Liberato, Tredicucci, Ciuti, Biasiol, Sorba, and
  Beltram}}]{anappara_signatures_2009}
\bibinfo{author}{\bibfnamefont{A.~A.} \bibnamefont{Anappara}},
  \bibinfo{author}{\bibfnamefont{S.}~\bibnamefont{De~Liberato}},
  \bibinfo{author}{\bibfnamefont{A.}~\bibnamefont{Tredicucci}},
  \bibinfo{author}{\bibfnamefont{C.}~\bibnamefont{Ciuti}},
  \bibinfo{author}{\bibfnamefont{G.}~\bibnamefont{Biasiol}},
  \bibinfo{author}{\bibfnamefont{L.}~\bibnamefont{Sorba}}, \bibnamefont{and}
  \bibinfo{author}{\bibfnamefont{F.}~\bibnamefont{Beltram}},
  \bibinfo{journal}{Physical Review B} \textbf{\bibinfo{volume}{79}},
  \bibinfo{pages}{201303} (\bibinfo{year}{2009}), \bibinfo{note}{publisher:
  American Physical Society},
  \urlprefix\url{https://link.aps.org/doi/10.1103/PhysRevB.79.201303}.

\bibitem[{\citenamefont{Ashhab and Nori}(2010)}]{ashhab_qubit-oscillator_2010}
\bibinfo{author}{\bibfnamefont{S.}~\bibnamefont{Ashhab}} \bibnamefont{and}
  \bibinfo{author}{\bibfnamefont{F.}~\bibnamefont{Nori}},
  \bibinfo{journal}{Physical Review A} \textbf{\bibinfo{volume}{81}},
  \bibinfo{pages}{042311} (\bibinfo{year}{2010}), ISSN
  \bibinfo{issn}{1050-2947, 1094-1622},
  \urlprefix\url{https://link.aps.org/doi/10.1103/PhysRevA.81.042311}.

\bibitem[{\citenamefont{Galego et~al.}(2015)\citenamefont{Galego, Garcia-Vidal,
  and Feist}}]{galego_cavity-induced_2015}
\bibinfo{author}{\bibfnamefont{J.}~\bibnamefont{Galego}},
  \bibinfo{author}{\bibfnamefont{F.~J.} \bibnamefont{Garcia-Vidal}},
  \bibnamefont{and} \bibinfo{author}{\bibfnamefont{J.}~\bibnamefont{Feist}},
  \bibinfo{journal}{Physical Review X} \textbf{\bibinfo{volume}{5}}
  (\bibinfo{year}{2015}).

\bibitem[{\citenamefont{Orgiu et~al.}(2015)\citenamefont{Orgiu, George,
  Hutchison, Devaux, Dayen, Doudin, Stellacci, Genet, Schachenmayer, Genes
  et~al.}}]{orgiu_conductivity_2015}
\bibinfo{author}{\bibfnamefont{E.}~\bibnamefont{Orgiu}},
  \bibinfo{author}{\bibfnamefont{J.}~\bibnamefont{George}},
  \bibinfo{author}{\bibfnamefont{J.~A.} \bibnamefont{Hutchison}},
  \bibinfo{author}{\bibfnamefont{E.}~\bibnamefont{Devaux}},
  \bibinfo{author}{\bibfnamefont{J.~F.} \bibnamefont{Dayen}},
  \bibinfo{author}{\bibfnamefont{B.}~\bibnamefont{Doudin}},
  \bibinfo{author}{\bibfnamefont{F.}~\bibnamefont{Stellacci}},
  \bibinfo{author}{\bibfnamefont{C.}~\bibnamefont{Genet}},
  \bibinfo{author}{\bibfnamefont{J.}~\bibnamefont{Schachenmayer}},
  \bibinfo{author}{\bibfnamefont{C.}~\bibnamefont{Genes}},
  \bibnamefont{et~al.}, \bibinfo{journal}{Nature Materials}
  \textbf{\bibinfo{volume}{14}}, \bibinfo{pages}{1123} (\bibinfo{year}{2015}),
  \urlprefix\url{https://doi.org/10.1038/nmat4392}.

\bibitem[{\citenamefont{Herrera and
  Spano}(2016)}]{herrera_cavity-controlled_2016}
\bibinfo{author}{\bibfnamefont{F.}~\bibnamefont{Herrera}} \bibnamefont{and}
  \bibinfo{author}{\bibfnamefont{F.~C.} \bibnamefont{Spano}},
  \bibinfo{journal}{Physical Review Letters} \textbf{\bibinfo{volume}{116}},
  \bibinfo{pages}{238301} (\bibinfo{year}{2016}), ISSN
  \bibinfo{issn}{0031-9007, 1079-7114},
  \urlprefix\url{https://link.aps.org/doi/10.1103/PhysRevLett.116.238301}.

\bibitem[{\citenamefont{Kockum et~al.}(2017{\natexlab{a}})\citenamefont{Kockum,
  Miranowicz, Macrì, Savasta, and Nori}}]{kockum_deterministic_2017}
\bibinfo{author}{\bibfnamefont{A.~F.} \bibnamefont{Kockum}},
  \bibinfo{author}{\bibfnamefont{A.}~\bibnamefont{Miranowicz}},
  \bibinfo{author}{\bibfnamefont{V.}~\bibnamefont{Macrì}},
  \bibinfo{author}{\bibfnamefont{S.}~\bibnamefont{Savasta}}, \bibnamefont{and}
  \bibinfo{author}{\bibfnamefont{F.}~\bibnamefont{Nori}},
  \bibinfo{journal}{Physical Review A} \textbf{\bibinfo{volume}{95}},
  \bibinfo{pages}{063849} (\bibinfo{year}{2017}{\natexlab{a}}), ISSN
  \bibinfo{issn}{2469-9926, 2469-9934},
  \urlprefix\url{http://link.aps.org/doi/10.1103/PhysRevA.95.063849}.

\bibitem[{\citenamefont{Hutchison et~al.}(2012)\citenamefont{Hutchison,
  Schwartz, Genet, Devaux, and Ebbesen}}]{hutchison_modifying_2012}
\bibinfo{author}{\bibfnamefont{J.~A.} \bibnamefont{Hutchison}},
  \bibinfo{author}{\bibfnamefont{T.}~\bibnamefont{Schwartz}},
  \bibinfo{author}{\bibfnamefont{C.}~\bibnamefont{Genet}},
  \bibinfo{author}{\bibfnamefont{E.}~\bibnamefont{Devaux}}, \bibnamefont{and}
  \bibinfo{author}{\bibfnamefont{T.~W.} \bibnamefont{Ebbesen}},
  \bibinfo{journal}{Angewandte Chemie (International Ed. in English)}
  \textbf{\bibinfo{volume}{51}}, \bibinfo{pages}{1592} (\bibinfo{year}{2012}),
  ISSN \bibinfo{issn}{1521-3773}.

\bibitem[{\citenamefont{Thomas et~al.}(2019)\citenamefont{Thomas, Devaux,
  Nagarajan, Chervy, Seidel, Hagenmüller, Schütz, Schachenmayer, Genet,
  Pupillo et~al.}}]{thomas_exploring_2019}
\bibinfo{author}{\bibfnamefont{A.}~\bibnamefont{Thomas}},
  \bibinfo{author}{\bibfnamefont{E.}~\bibnamefont{Devaux}},
  \bibinfo{author}{\bibfnamefont{K.}~\bibnamefont{Nagarajan}},
  \bibinfo{author}{\bibfnamefont{T.}~\bibnamefont{Chervy}},
  \bibinfo{author}{\bibfnamefont{M.}~\bibnamefont{Seidel}},
  \bibinfo{author}{\bibfnamefont{D.}~\bibnamefont{Hagenmüller}},
  \bibinfo{author}{\bibfnamefont{S.}~\bibnamefont{Schütz}},
  \bibinfo{author}{\bibfnamefont{J.}~\bibnamefont{Schachenmayer}},
  \bibinfo{author}{\bibfnamefont{C.}~\bibnamefont{Genet}},
  \bibinfo{author}{\bibfnamefont{G.}~\bibnamefont{Pupillo}},
  \bibnamefont{et~al.}, \bibinfo{journal}{arXiv:1911.01459 [cond-mat,
  physics:quant-ph]}  (\bibinfo{year}{2019}), \bibinfo{note}{arXiv:
  1911.01459}, \urlprefix\url{http://arxiv.org/abs/1911.01459}.

\bibitem[{\citenamefont{Paravicini-Bagliani
  et~al.}(2019)\citenamefont{Paravicini-Bagliani, Appugliese, Richter,
  Valmorra, Keller, Beck, Bartolo, Rössler, Ihn, Ensslin
  et~al.}}]{paravicini-bagliani_magneto-transport_2019}
\bibinfo{author}{\bibfnamefont{G.~L.} \bibnamefont{Paravicini-Bagliani}},
  \bibinfo{author}{\bibfnamefont{F.}~\bibnamefont{Appugliese}},
  \bibinfo{author}{\bibfnamefont{E.}~\bibnamefont{Richter}},
  \bibinfo{author}{\bibfnamefont{F.}~\bibnamefont{Valmorra}},
  \bibinfo{author}{\bibfnamefont{J.}~\bibnamefont{Keller}},
  \bibinfo{author}{\bibfnamefont{M.}~\bibnamefont{Beck}},
  \bibinfo{author}{\bibfnamefont{N.}~\bibnamefont{Bartolo}},
  \bibinfo{author}{\bibfnamefont{C.}~\bibnamefont{Rössler}},
  \bibinfo{author}{\bibfnamefont{T.}~\bibnamefont{Ihn}},
  \bibinfo{author}{\bibfnamefont{K.}~\bibnamefont{Ensslin}},
  \bibnamefont{et~al.}, \bibinfo{journal}{Nature Physics}
  \textbf{\bibinfo{volume}{15}}, \bibinfo{pages}{186} (\bibinfo{year}{2019}),
  ISSN \bibinfo{issn}{1745-2481}, \bibinfo{note}{number: 2 Publisher: Nature
  Publishing Group},
  \urlprefix\url{https://www.nature.com/articles/s41567-018-0346-y}.

\bibitem[{\citenamefont{Gambino et~al.}(2014)\citenamefont{Gambino, Mazzeo,
  Genco, Di~Stefano, Savasta, Patanè, Ballarini, Mangione, Lerario, Sanvitto
  et~al.}}]{gambino_exploring_2014}
\bibinfo{author}{\bibfnamefont{S.}~\bibnamefont{Gambino}},
  \bibinfo{author}{\bibfnamefont{M.}~\bibnamefont{Mazzeo}},
  \bibinfo{author}{\bibfnamefont{A.}~\bibnamefont{Genco}},
  \bibinfo{author}{\bibfnamefont{O.}~\bibnamefont{Di~Stefano}},
  \bibinfo{author}{\bibfnamefont{S.}~\bibnamefont{Savasta}},
  \bibinfo{author}{\bibfnamefont{S.}~\bibnamefont{Patanè}},
  \bibinfo{author}{\bibfnamefont{D.}~\bibnamefont{Ballarini}},
  \bibinfo{author}{\bibfnamefont{F.}~\bibnamefont{Mangione}},
  \bibinfo{author}{\bibfnamefont{G.}~\bibnamefont{Lerario}},
  \bibinfo{author}{\bibfnamefont{D.}~\bibnamefont{Sanvitto}},
  \bibnamefont{et~al.}, \bibinfo{journal}{ACS Photonics}
  \textbf{\bibinfo{volume}{1}}, \bibinfo{pages}{1042} (\bibinfo{year}{2014}),
  \bibinfo{note}{publisher: American Chemical Society},
  \urlprefix\url{https://doi.org/10.1021/ph500266d}.

\bibitem[{\citenamefont{Todisco et~al.}(2018)\citenamefont{Todisco, De~Giorgi,
  Esposito, De~Marco, Zizzari, Bianco, Dominici, Ballarini, Arima, Gigli
  et~al.}}]{todisco_ultrastrong_2018}
\bibinfo{author}{\bibfnamefont{F.}~\bibnamefont{Todisco}},
  \bibinfo{author}{\bibfnamefont{M.}~\bibnamefont{De~Giorgi}},
  \bibinfo{author}{\bibfnamefont{M.}~\bibnamefont{Esposito}},
  \bibinfo{author}{\bibfnamefont{L.}~\bibnamefont{De~Marco}},
  \bibinfo{author}{\bibfnamefont{A.}~\bibnamefont{Zizzari}},
  \bibinfo{author}{\bibfnamefont{M.}~\bibnamefont{Bianco}},
  \bibinfo{author}{\bibfnamefont{L.}~\bibnamefont{Dominici}},
  \bibinfo{author}{\bibfnamefont{D.}~\bibnamefont{Ballarini}},
  \bibinfo{author}{\bibfnamefont{V.}~\bibnamefont{Arima}},
  \bibinfo{author}{\bibfnamefont{G.}~\bibnamefont{Gigli}},
  \bibnamefont{et~al.}, \bibinfo{journal}{ACS Photonics}
  \textbf{\bibinfo{volume}{5}}, \bibinfo{pages}{143} (\bibinfo{year}{2018}),
  \bibinfo{note}{publisher: American Chemical Society},
  \urlprefix\url{https://doi.org/10.1021/acsphotonics.7b00554}.

\bibitem[{\citenamefont{Mazzeo et~al.}(2014)\citenamefont{Mazzeo, Genco,
  Gambino, Ballarini, Mangione, Di~Stefano, Patanè, Savasta, Sanvitto, and
  Gigli}}]{mazzeo_ultrastrong_2014}
\bibinfo{author}{\bibfnamefont{M.}~\bibnamefont{Mazzeo}},
  \bibinfo{author}{\bibfnamefont{A.}~\bibnamefont{Genco}},
  \bibinfo{author}{\bibfnamefont{S.}~\bibnamefont{Gambino}},
  \bibinfo{author}{\bibfnamefont{D.}~\bibnamefont{Ballarini}},
  \bibinfo{author}{\bibfnamefont{F.}~\bibnamefont{Mangione}},
  \bibinfo{author}{\bibfnamefont{O.}~\bibnamefont{Di~Stefano}},
  \bibinfo{author}{\bibfnamefont{S.}~\bibnamefont{Patanè}},
  \bibinfo{author}{\bibfnamefont{S.}~\bibnamefont{Savasta}},
  \bibinfo{author}{\bibfnamefont{D.}~\bibnamefont{Sanvitto}}, \bibnamefont{and}
  \bibinfo{author}{\bibfnamefont{G.}~\bibnamefont{Gigli}},
  \bibinfo{journal}{Applied Physics Letters} \textbf{\bibinfo{volume}{104}},
  \bibinfo{pages}{233303} (\bibinfo{year}{2014}), ISSN
  \bibinfo{issn}{0003-6951}, \bibinfo{note}{publisher: American Institute of
  Physics}, \urlprefix\url{https://aip.scitation.org/doi/10.1063/1.4882422}.

\bibitem[{\citenamefont{Günter et~al.}(2009)\citenamefont{Günter, Anappara,
  Hees, Sell, Biasiol, Sorba, De~Liberato, Ciuti, Tredicucci, Leitenstorfer
  et~al.}}]{gunter_sub-cycle_2009}
\bibinfo{author}{\bibfnamefont{G.}~\bibnamefont{Günter}},
  \bibinfo{author}{\bibfnamefont{A.~A.} \bibnamefont{Anappara}},
  \bibinfo{author}{\bibfnamefont{J.}~\bibnamefont{Hees}},
  \bibinfo{author}{\bibfnamefont{A.}~\bibnamefont{Sell}},
  \bibinfo{author}{\bibfnamefont{G.}~\bibnamefont{Biasiol}},
  \bibinfo{author}{\bibfnamefont{L.}~\bibnamefont{Sorba}},
  \bibinfo{author}{\bibfnamefont{S.}~\bibnamefont{De~Liberato}},
  \bibinfo{author}{\bibfnamefont{C.}~\bibnamefont{Ciuti}},
  \bibinfo{author}{\bibfnamefont{A.}~\bibnamefont{Tredicucci}},
  \bibinfo{author}{\bibfnamefont{A.}~\bibnamefont{Leitenstorfer}},
  \bibnamefont{et~al.}, \bibinfo{journal}{Nature}
  \textbf{\bibinfo{volume}{458}}, \bibinfo{pages}{178} (\bibinfo{year}{2009}),
  ISSN \bibinfo{issn}{1476-4687}, \bibinfo{note}{number: 7235 Publisher: Nature
  Publishing Group},
  \urlprefix\url{http://www.nature.com/articles/nature07838}.

\bibitem[{\citenamefont{Delteil et~al.}(2012)\citenamefont{Delteil, Vasanelli,
  Todorov, Feuillet~Palma, Renaudat St-Jean, Beaudoin, Sagnes, and
  Sirtori}}]{delteil_charge-induced_2012}
\bibinfo{author}{\bibfnamefont{A.}~\bibnamefont{Delteil}},
  \bibinfo{author}{\bibfnamefont{A.}~\bibnamefont{Vasanelli}},
  \bibinfo{author}{\bibfnamefont{Y.}~\bibnamefont{Todorov}},
  \bibinfo{author}{\bibfnamefont{C.}~\bibnamefont{Feuillet~Palma}},
  \bibinfo{author}{\bibfnamefont{M.}~\bibnamefont{Renaudat St-Jean}},
  \bibinfo{author}{\bibfnamefont{G.}~\bibnamefont{Beaudoin}},
  \bibinfo{author}{\bibfnamefont{I.}~\bibnamefont{Sagnes}}, \bibnamefont{and}
  \bibinfo{author}{\bibfnamefont{C.}~\bibnamefont{Sirtori}},
  \bibinfo{journal}{Physical Review Letters} \textbf{\bibinfo{volume}{109}},
  \bibinfo{pages}{246808} (\bibinfo{year}{2012}), \bibinfo{note}{publisher:
  American Physical Society},
  \urlprefix\url{https://link.aps.org/doi/10.1103/PhysRevLett.109.246808}.

\bibitem[{\citenamefont{Auer and Burkard}(2012)}]{auer_entangled_2012}
\bibinfo{author}{\bibfnamefont{A.}~\bibnamefont{Auer}} \bibnamefont{and}
  \bibinfo{author}{\bibfnamefont{G.}~\bibnamefont{Burkard}},
  \bibinfo{journal}{Physical Review B} \textbf{\bibinfo{volume}{85}},
  \bibinfo{pages}{235140} (\bibinfo{year}{2012}), \bibinfo{note}{publisher:
  American Physical Society},
  \urlprefix\url{https://link.aps.org/doi/10.1103/PhysRevB.85.235140}.

\bibitem[{\citenamefont{Niemczyk et~al.}(2010)\citenamefont{Niemczyk, Deppe,
  Huebl, Menzel, Hocke, Schwarz, Garcia-Ripoll, Zueco, Hümmer, Solano
  et~al.}}]{niemczyk_circuit_2010}
\bibinfo{author}{\bibfnamefont{T.}~\bibnamefont{Niemczyk}},
  \bibinfo{author}{\bibfnamefont{F.}~\bibnamefont{Deppe}},
  \bibinfo{author}{\bibfnamefont{H.}~\bibnamefont{Huebl}},
  \bibinfo{author}{\bibfnamefont{E.~P.} \bibnamefont{Menzel}},
  \bibinfo{author}{\bibfnamefont{F.}~\bibnamefont{Hocke}},
  \bibinfo{author}{\bibfnamefont{M.~J.} \bibnamefont{Schwarz}},
  \bibinfo{author}{\bibfnamefont{J.~J.} \bibnamefont{Garcia-Ripoll}},
  \bibinfo{author}{\bibfnamefont{D.}~\bibnamefont{Zueco}},
  \bibinfo{author}{\bibfnamefont{T.}~\bibnamefont{Hümmer}},
  \bibinfo{author}{\bibfnamefont{E.}~\bibnamefont{Solano}},
  \bibnamefont{et~al.}, \bibinfo{journal}{Nature Physics}
  \textbf{\bibinfo{volume}{6}}, \bibinfo{pages}{772} (\bibinfo{year}{2010}),
  ISSN \bibinfo{issn}{1745-2481}, \bibinfo{note}{number: 10 Publisher: Nature
  Publishing Group}, \urlprefix\url{http://www.nature.com/articles/nphys1730}.

\bibitem[{\citenamefont{Forn-Díaz et~al.}(2010)\citenamefont{Forn-Díaz,
  Lisenfeld, Marcos, García-Ripoll, Solano, Harmans, and
  Mooij}}]{forn-diaz_observation_2010}
\bibinfo{author}{\bibfnamefont{P.}~\bibnamefont{Forn-Díaz}},
  \bibinfo{author}{\bibfnamefont{J.}~\bibnamefont{Lisenfeld}},
  \bibinfo{author}{\bibfnamefont{D.}~\bibnamefont{Marcos}},
  \bibinfo{author}{\bibfnamefont{J.~J.} \bibnamefont{García-Ripoll}},
  \bibinfo{author}{\bibfnamefont{E.}~\bibnamefont{Solano}},
  \bibinfo{author}{\bibfnamefont{C.~J. P.~M.} \bibnamefont{Harmans}},
  \bibnamefont{and} \bibinfo{author}{\bibfnamefont{J.~E.} \bibnamefont{Mooij}},
  \bibinfo{journal}{Physical Review Letters} \textbf{\bibinfo{volume}{105}},
  \bibinfo{pages}{237001} (\bibinfo{year}{2010}), ISSN
  \bibinfo{issn}{0031-9007, 1079-7114},
  \urlprefix\url{https://link.aps.org/doi/10.1103/PhysRevLett.105.237001}.

\bibitem[{\citenamefont{Forn-Díaz et~al.}(2017)\citenamefont{Forn-Díaz,
  García-Ripoll, Peropadre, Orgiazzi, Yurtalan, Belyansky, Wilson, and
  Lupascu}}]{forn-diaz_ultrastrong_2017}
\bibinfo{author}{\bibfnamefont{P.}~\bibnamefont{Forn-Díaz}},
  \bibinfo{author}{\bibfnamefont{J.~J.} \bibnamefont{García-Ripoll}},
  \bibinfo{author}{\bibfnamefont{B.}~\bibnamefont{Peropadre}},
  \bibinfo{author}{\bibfnamefont{J.-L.} \bibnamefont{Orgiazzi}},
  \bibinfo{author}{\bibfnamefont{M.~A.} \bibnamefont{Yurtalan}},
  \bibinfo{author}{\bibfnamefont{R.}~\bibnamefont{Belyansky}},
  \bibinfo{author}{\bibfnamefont{C.~M.} \bibnamefont{Wilson}},
  \bibnamefont{and} \bibinfo{author}{\bibfnamefont{A.}~\bibnamefont{Lupascu}},
  \bibinfo{journal}{Nature Physics} \textbf{\bibinfo{volume}{13}},
  \bibinfo{pages}{39} (\bibinfo{year}{2017}), ISSN \bibinfo{issn}{1745-2481},
  \bibinfo{note}{number: 1 Publisher: Nature Publishing Group},
  \urlprefix\url{http://www.nature.com/articles/nphys3905}.

\bibitem[{\citenamefont{Puertas~Martínez
  et~al.}(2019)\citenamefont{Puertas~Martínez, Léger, Gheeraert,
  Dassonneville, Planat, Foroughi, Krupko, Buisson, Naud, Hasch-Guichard
  et~al.}}]{puertas_martinez_tunable_2019}
\bibinfo{author}{\bibfnamefont{J.}~\bibnamefont{Puertas~Martínez}},
  \bibinfo{author}{\bibfnamefont{S.}~\bibnamefont{Léger}},
  \bibinfo{author}{\bibfnamefont{N.}~\bibnamefont{Gheeraert}},
  \bibinfo{author}{\bibfnamefont{R.}~\bibnamefont{Dassonneville}},
  \bibinfo{author}{\bibfnamefont{L.}~\bibnamefont{Planat}},
  \bibinfo{author}{\bibfnamefont{F.}~\bibnamefont{Foroughi}},
  \bibinfo{author}{\bibfnamefont{Y.}~\bibnamefont{Krupko}},
  \bibinfo{author}{\bibfnamefont{O.}~\bibnamefont{Buisson}},
  \bibinfo{author}{\bibfnamefont{C.}~\bibnamefont{Naud}},
  \bibinfo{author}{\bibfnamefont{W.}~\bibnamefont{Hasch-Guichard}},
  \bibnamefont{et~al.}, \bibinfo{journal}{npj Quantum Information}
  \textbf{\bibinfo{volume}{5}}, \bibinfo{pages}{1} (\bibinfo{year}{2019}), ISSN
  \bibinfo{issn}{2056-6387}, \bibinfo{note}{number: 1 Publisher: Nature
  Publishing Group},
  \urlprefix\url{http://www.nature.com/articles/s41534-018-0104-0}.

\bibitem[{\citenamefont{Magazzù et~al.}(2018)\citenamefont{Magazzù,
  Forn-Díaz, Belyansky, Orgiazzi, Yurtalan, Otto, Lupascu, Wilson, and
  Grifoni}}]{magazzu_probing_2018}
\bibinfo{author}{\bibfnamefont{L.}~\bibnamefont{Magazzù}},
  \bibinfo{author}{\bibfnamefont{P.}~\bibnamefont{Forn-Díaz}},
  \bibinfo{author}{\bibfnamefont{R.}~\bibnamefont{Belyansky}},
  \bibinfo{author}{\bibfnamefont{J.-L.} \bibnamefont{Orgiazzi}},
  \bibinfo{author}{\bibfnamefont{M.~A.} \bibnamefont{Yurtalan}},
  \bibinfo{author}{\bibfnamefont{M.~R.} \bibnamefont{Otto}},
  \bibinfo{author}{\bibfnamefont{A.}~\bibnamefont{Lupascu}},
  \bibinfo{author}{\bibfnamefont{C.~M.} \bibnamefont{Wilson}},
  \bibnamefont{and} \bibinfo{author}{\bibfnamefont{M.}~\bibnamefont{Grifoni}},
  \bibinfo{journal}{Nature Communications} \textbf{\bibinfo{volume}{9}},
  \bibinfo{pages}{1403} (\bibinfo{year}{2018}), ISSN \bibinfo{issn}{2041-1723},
  \bibinfo{note}{number: 1 Publisher: Nature Publishing Group},
  \urlprefix\url{http://www.nature.com/articles/s41467-018-03626-w}.

\bibitem[{\citenamefont{Chikkaraddy et~al.}(2016)\citenamefont{Chikkaraddy,
  de~Nijs, Benz, Barrow, Scherman, Rosta, Demetriadou, Fox, Hess, and
  Baumberg}}]{chikkaraddy_single-molecule_2016}
\bibinfo{author}{\bibfnamefont{R.}~\bibnamefont{Chikkaraddy}},
  \bibinfo{author}{\bibfnamefont{B.}~\bibnamefont{de~Nijs}},
  \bibinfo{author}{\bibfnamefont{F.}~\bibnamefont{Benz}},
  \bibinfo{author}{\bibfnamefont{S.~J.} \bibnamefont{Barrow}},
  \bibinfo{author}{\bibfnamefont{O.~A.} \bibnamefont{Scherman}},
  \bibinfo{author}{\bibfnamefont{E.}~\bibnamefont{Rosta}},
  \bibinfo{author}{\bibfnamefont{A.}~\bibnamefont{Demetriadou}},
  \bibinfo{author}{\bibfnamefont{P.}~\bibnamefont{Fox}},
  \bibinfo{author}{\bibfnamefont{O.}~\bibnamefont{Hess}}, \bibnamefont{and}
  \bibinfo{author}{\bibfnamefont{J.~J.} \bibnamefont{Baumberg}},
  \bibinfo{journal}{Nature; London} \textbf{\bibinfo{volume}{535}},
  \bibinfo{pages}{127} (\bibinfo{year}{2016}), ISSN \bibinfo{issn}{00280836},
  \bibinfo{note}{num Pages: 4 Place: London, United States, London Publisher:
  Nature Publishing Group Section: LETTER},
  \urlprefix\url{http://search.proquest.com/docview/1809935134/abstract/3830216F670D43B9PQ/1}.

\bibitem[{\citenamefont{Frisk~Kockum et~al.}(2019)\citenamefont{Frisk~Kockum,
  Miranowicz, De~Liberato, Savasta, and Nori}}]{frisk_kockum_ultrastrong_2019}
\bibinfo{author}{\bibfnamefont{A.}~\bibnamefont{Frisk~Kockum}},
  \bibinfo{author}{\bibfnamefont{A.}~\bibnamefont{Miranowicz}},
  \bibinfo{author}{\bibfnamefont{S.}~\bibnamefont{De~Liberato}},
  \bibinfo{author}{\bibfnamefont{S.}~\bibnamefont{Savasta}}, \bibnamefont{and}
  \bibinfo{author}{\bibfnamefont{F.}~\bibnamefont{Nori}},
  \bibinfo{journal}{Nature Reviews Physics} \textbf{\bibinfo{volume}{1}},
  \bibinfo{pages}{19} (\bibinfo{year}{2019}), ISSN \bibinfo{issn}{2522-5820},
  \bibinfo{note}{number: 1 Publisher: Nature Publishing Group},
  \urlprefix\url{http://www.nature.com/articles/s42254-018-0006-2}.

\bibitem[{\citenamefont{Purcell}(1995)}]{purcell_spontaneous_1995}
\bibinfo{author}{\bibfnamefont{E.~M.} \bibnamefont{Purcell}}, in
  \emph{\bibinfo{booktitle}{Confined {Electrons} and {Photons}: {New} {Physics}
  and {Applications}}}, edited by
  \bibinfo{editor}{\bibfnamefont{E.}~\bibnamefont{Burstein}} \bibnamefont{and}
  \bibinfo{editor}{\bibfnamefont{C.}~\bibnamefont{Weisbuch}}
  (\bibinfo{publisher}{Springer US}, \bibinfo{address}{Boston, MA},
  \bibinfo{year}{1995}), {NATO} {ASI} {Series}, pp. \bibinfo{pages}{839--839},
  ISBN \bibinfo{isbn}{978-1-4615-1963-8},
  \urlprefix\url{https://doi.org/10.1007/978-1-4615-1963-8_40}.

\bibitem[{\citenamefont{Hines et~al.}(2004)\citenamefont{Hines, Dawson,
  McKenzie, and Milburn}}]{hines_entanglement_2004}
\bibinfo{author}{\bibfnamefont{A.~P.} \bibnamefont{Hines}},
  \bibinfo{author}{\bibfnamefont{C.~M.} \bibnamefont{Dawson}},
  \bibinfo{author}{\bibfnamefont{R.~H.} \bibnamefont{McKenzie}},
  \bibnamefont{and} \bibinfo{author}{\bibfnamefont{G.~J.}
  \bibnamefont{Milburn}}, \bibinfo{journal}{Physical Review A}
  \textbf{\bibinfo{volume}{70}}, \bibinfo{pages}{022303}
  (\bibinfo{year}{2004}), \bibinfo{note}{publisher: American Physical Society},
  \urlprefix\url{https://link.aps.org/doi/10.1103/PhysRevA.70.022303}.

\bibitem[{\citenamefont{Beaudoin et~al.}(2011)\citenamefont{Beaudoin, Gambetta,
  and Blais}}]{beaudoin_dissipation_2011}
\bibinfo{author}{\bibfnamefont{F.}~\bibnamefont{Beaudoin}},
  \bibinfo{author}{\bibfnamefont{J.~M.} \bibnamefont{Gambetta}},
  \bibnamefont{and} \bibinfo{author}{\bibfnamefont{A.}~\bibnamefont{Blais}},
  \bibinfo{journal}{Physical Review A} \textbf{\bibinfo{volume}{84}},
  \bibinfo{pages}{043832} (\bibinfo{year}{2011}), ISSN
  \bibinfo{issn}{1050-2947, 1094-1622},
  \urlprefix\url{https://link.aps.org/doi/10.1103/PhysRevA.84.043832}.

\bibitem[{\citenamefont{Jaynes and Cummings}(1963)}]{jaynes_comparison_1963}
\bibinfo{author}{\bibfnamefont{E.}~\bibnamefont{Jaynes}} \bibnamefont{and}
  \bibinfo{author}{\bibfnamefont{F.}~\bibnamefont{Cummings}},
  \bibinfo{journal}{Proceedings of the IEEE} \textbf{\bibinfo{volume}{51}},
  \bibinfo{pages}{89} (\bibinfo{year}{1963}), ISSN \bibinfo{issn}{1558-2256},
  \bibinfo{note}{conference Name: Proceedings of the IEEE}.

\bibitem[{\citenamefont{Krantz et~al.}(2019)\citenamefont{Krantz, Kjaergaard,
  Yan, Orlando, Gustavsson, and Oliver}}]{krantz_quantum_2019}
\bibinfo{author}{\bibfnamefont{P.}~\bibnamefont{Krantz}},
  \bibinfo{author}{\bibfnamefont{M.}~\bibnamefont{Kjaergaard}},
  \bibinfo{author}{\bibfnamefont{F.}~\bibnamefont{Yan}},
  \bibinfo{author}{\bibfnamefont{T.~P.} \bibnamefont{Orlando}},
  \bibinfo{author}{\bibfnamefont{S.}~\bibnamefont{Gustavsson}},
  \bibnamefont{and} \bibinfo{author}{\bibfnamefont{W.~D.}
  \bibnamefont{Oliver}}, \bibinfo{journal}{Applied Physics Reviews}
  \textbf{\bibinfo{volume}{6}}, \bibinfo{pages}{021318} (\bibinfo{year}{2019}),
  ISSN \bibinfo{issn}{1931-9401},
  \urlprefix\url{http://aip.scitation.org/doi/10.1063/1.5089550}.

\bibitem[{\citenamefont{Ozfidan et~al.}(2020)\citenamefont{Ozfidan, Deng,
  Smirnov, Lanting, Harris, Swenson, Whittaker, Altomare, Babcock, Baron
  et~al.}}]{ozfidan_demonstration_2020}
\bibinfo{author}{\bibfnamefont{I.}~\bibnamefont{Ozfidan}},
  \bibinfo{author}{\bibfnamefont{C.}~\bibnamefont{Deng}},
  \bibinfo{author}{\bibfnamefont{A.}~\bibnamefont{Smirnov}},
  \bibinfo{author}{\bibfnamefont{T.}~\bibnamefont{Lanting}},
  \bibinfo{author}{\bibfnamefont{R.}~\bibnamefont{Harris}},
  \bibinfo{author}{\bibfnamefont{L.}~\bibnamefont{Swenson}},
  \bibinfo{author}{\bibfnamefont{J.}~\bibnamefont{Whittaker}},
  \bibinfo{author}{\bibfnamefont{F.}~\bibnamefont{Altomare}},
  \bibinfo{author}{\bibfnamefont{M.}~\bibnamefont{Babcock}},
  \bibinfo{author}{\bibfnamefont{C.}~\bibnamefont{Baron}},
  \bibnamefont{et~al.}, \bibinfo{journal}{Physical Review Applied}
  \textbf{\bibinfo{volume}{13}}, \bibinfo{pages}{034037}
  (\bibinfo{year}{2020}), ISSN \bibinfo{issn}{2331-7019},
  \urlprefix\url{https://link.aps.org/doi/10.1103/PhysRevApplied.13.034037}.

\bibitem[{\citenamefont{Yan et~al.}(2016)\citenamefont{Yan, Gustavsson, Kamal,
  Birenbaum, Sears, Hover, Gudmundsen, Rosenberg, Samach, Weber
  et~al.}}]{yan_flux_2016}
\bibinfo{author}{\bibfnamefont{F.}~\bibnamefont{Yan}},
  \bibinfo{author}{\bibfnamefont{S.}~\bibnamefont{Gustavsson}},
  \bibinfo{author}{\bibfnamefont{A.}~\bibnamefont{Kamal}},
  \bibinfo{author}{\bibfnamefont{J.}~\bibnamefont{Birenbaum}},
  \bibinfo{author}{\bibfnamefont{A.~P.} \bibnamefont{Sears}},
  \bibinfo{author}{\bibfnamefont{D.}~\bibnamefont{Hover}},
  \bibinfo{author}{\bibfnamefont{T.~J.} \bibnamefont{Gudmundsen}},
  \bibinfo{author}{\bibfnamefont{D.}~\bibnamefont{Rosenberg}},
  \bibinfo{author}{\bibfnamefont{G.}~\bibnamefont{Samach}},
  \bibinfo{author}{\bibfnamefont{S.}~\bibnamefont{Weber}},
  \bibnamefont{et~al.}, \bibinfo{journal}{Nature Communications}
  \textbf{\bibinfo{volume}{7}}, \bibinfo{pages}{12964} (\bibinfo{year}{2016}),
  ISSN \bibinfo{issn}{2041-1723}, \bibinfo{note}{bandiera\_abtest: a
  Cc\_license\_type: cc\_by Cg\_type: Nature Research Journals Number: 1
  Primary\_atype: Research Publisher: Nature Publishing Group Subject\_term:
  Electrical and electronic engineering;Quantum information;Qubits
  Subject\_term\_id:
  electrical-and-electronic-engineering;quantum-information;qubits},
  \urlprefix\url{https://www.nature.com/articles/ncomms12964}.

\bibitem[{\citenamefont{Chong et~al.}(2007)\citenamefont{Chong, Pritchard, and
  Solja{\v c}i{\'c}}}]{Chong2007}
\bibinfo{author}{\bibfnamefont{Y.~D.} \bibnamefont{Chong}},
  \bibinfo{author}{\bibfnamefont{D.~E.} \bibnamefont{Pritchard}},
  \bibnamefont{and} \bibinfo{author}{\bibfnamefont{M.}~\bibnamefont{Solja{\v
  c}i{\'c}}}, \bibinfo{journal}{Physical Review B}
  \textbf{\bibinfo{volume}{75}}, \bibinfo{pages}{235124}
  (\bibinfo{year}{2007}), ISSN \bibinfo{issn}{1098-0121, 1550-235X}.

\bibitem[{\citenamefont{Kurman and Kaminer}(2020)}]{kurman_tunable_2020}
\bibinfo{author}{\bibfnamefont{Y.}~\bibnamefont{Kurman}} \bibnamefont{and}
  \bibinfo{author}{\bibfnamefont{I.}~\bibnamefont{Kaminer}},
  \bibinfo{journal}{Nature Physics} pp. \bibinfo{pages}{1--7}
  (\bibinfo{year}{2020}), ISSN \bibinfo{issn}{1745-2481},
  \bibinfo{note}{publisher: Nature Publishing Group},
  \urlprefix\url{http://www.nature.com/articles/s41567-020-0890-0}.

\bibitem[{\citenamefont{Garziano et~al.}(2015)\citenamefont{Garziano, Stassi,
  Macrì, Kockum, Savasta, and Nori}}]{garziano_multiphoton_2015}
\bibinfo{author}{\bibfnamefont{L.}~\bibnamefont{Garziano}},
  \bibinfo{author}{\bibfnamefont{R.}~\bibnamefont{Stassi}},
  \bibinfo{author}{\bibfnamefont{V.}~\bibnamefont{Macrì}},
  \bibinfo{author}{\bibfnamefont{A.~F.} \bibnamefont{Kockum}},
  \bibinfo{author}{\bibfnamefont{S.}~\bibnamefont{Savasta}}, \bibnamefont{and}
  \bibinfo{author}{\bibfnamefont{F.}~\bibnamefont{Nori}},
  \bibinfo{journal}{Physical Review A} \textbf{\bibinfo{volume}{92}},
  \bibinfo{pages}{063830} (\bibinfo{year}{2015}), ISSN
  \bibinfo{issn}{1050-2947, 1094-1622},
  \urlprefix\url{https://link.aps.org/doi/10.1103/PhysRevA.92.063830}.

\bibitem[{\citenamefont{Kockum et~al.}(2017{\natexlab{b}})\citenamefont{Kockum,
  Macrì, Garziano, Savasta, and Nori}}]{kockum_frequency_2017}
\bibinfo{author}{\bibfnamefont{A.~F.} \bibnamefont{Kockum}},
  \bibinfo{author}{\bibfnamefont{V.}~\bibnamefont{Macrì}},
  \bibinfo{author}{\bibfnamefont{L.}~\bibnamefont{Garziano}},
  \bibinfo{author}{\bibfnamefont{S.}~\bibnamefont{Savasta}}, \bibnamefont{and}
  \bibinfo{author}{\bibfnamefont{F.}~\bibnamefont{Nori}},
  \bibinfo{journal}{Scientific Reports} \textbf{\bibinfo{volume}{7}},
  \bibinfo{pages}{5313} (\bibinfo{year}{2017}{\natexlab{b}}), ISSN
  \bibinfo{issn}{2045-2322}, \bibinfo{note}{number: 1 Publisher: Nature
  Publishing Group},
  \urlprefix\url{http://www.nature.com/articles/s41598-017-04225-3}.

\end{thebibliography}
\end{document}